\documentclass[aip,jap,amsmath,amssymb,reprint,floatfix]{revtex4-1}
\usepackage{graphicx}
\usepackage{hyperref}
\usepackage{dcolumn}
\usepackage{diagbox}

\begin{document}

\title{Simulation of the electrical conductivity of two-dimensional films with aligned rod-like conductive fillers: effect of the filler length dispersity}

\author{Yuri~Yu.~Tarasevich}
\email[Corresponding author: ]{tarasevich@asu.edu.ru}

\author{Irina~V.~Vodolazskaya}
\email{vodolazskaya\_agu@mail.ru}

\author{Andrei~V.~Eserkepov}
\email{dantealigjery49@gmail.com}

\author{Valeria~A.~Goltseva}
\email{valeria.lozunova@gmail.com}

\author{Petr~G.~Selin}
\email{selinpeter1997@mail.ru}
\affiliation{Laboratory of Mathematical Modeling, Astrakhan State University, Astrakhan, Russia, 414056}

\author{Nikolai~I.~Lebovka}
\email{lebovka@gmail.com}
\affiliation{Department of Physical Chemistry of Disperse Minerals, F.~D.~Ovcharenko Institute of Biocolloidal Chemistry, NAS of Ukraine, Kiev, Ukraine, 03142}
\affiliation{Department of Physics, Taras Shevchenko Kiev National University, Kiev, Ukraine, 01033}

\date{\today}

\begin{abstract}
Using Monte Carlo simulation, we studied the electrical conductivity of two-dimensional films. The films consisted of a poorly conductive host matrix and highly conductive rod-like fillers (rods). The rods were of various lengths fitting a log-normal distribution. They could be aligned along a direction. Special attention was paid to films having completely aligned rods. The impact of length dispersity and the extent of rod alignment on the insulator-to-conductor phase transition was studied. The greater the length dispersity the smaller the critical concentration. The anisotropy of the electrical conductivity was more pronounced in the vicinity of the phase transition. A finite size effect was found to be significant only in the vicinity of the phase transition.
\end{abstract}


\maketitle

\section{Introduction}\label{sec:intro}
Transparent conductive films containing highly conductive particles in a transparent, poorly conductive medium are important components of modern optoelectronic devices, for the improvement of which, numerous scientific studies are currently being carried out, including efforts to identify the main factors affecting the effective electrical conductivity and transparency of such films~\cite{Hecht2011AM}. Transparent conductors (TCs) are in demand in many modern devices such as touch screens, liquid crystal and organic light emitting displays, and solar cells. Advances in nanomaterial research have offered new TCs in place of the traditionally-used doped metal oxides. These TCs include carbon nanotubes (CNTs), graphene, metal nanowires (NWs), and printable metal grids.

Films with metallic NWs or CNTs, the aspect ratios of which (the ratio of the characteristic length to the characteristic transverse dimension of the particle) are of the order of hundreds and even thousands~\cite{Khanarian2013JAP,Borchert2015N,Forro2013AIPA}, show good conductivity and transparency.
There is a small amount of experimental data showing that electrical conductivity increases with increased average length of silver NWs~\cite{Bergin2012N,Sorel2012N,Langley2018NH}, as well as with an increase in the aspect ratios of silver NWs~\cite{White2010AFM} and of multiwalled CNTs~\cite{Gnanasekaran2014}.

The  length dependence of on-currents and the apparent mobilities for selectively dispersed, semiconducting single-walled CNTs (SWCNTs) has been investigated~\cite{Malhofer2017OE}. Transistors with the highest concentrations of long nanotubes approach the on-conductance and mobility values of pure semiconducting nanotube networks. The length and, in particular, the number, of the longest nanotubes in the composite play an important role in determining the final on-conductances and mobilities.

It should be noted that, in composites, the CNTs,  silver NWs and copper NWs differ not only in length but also in diameter; they may also have a curved shape; moreover, CNTs have dispersity in their specific conductivity. (According to IUPAC recommendations, we use the term ``dispersity'' here and below, no matter what term was used in the cited article, since ````Monodisperse'' is a self-contradictory term, and ``polydisperse'' is tautologous.''\cite{Stepto2009PAC}.)  All these factors affect the effective electrical conductivity of the composite and complicate analysis of the experimental data.
Statistical studies of the distribution of filler lengths have been carried out for copper NWs~\cite{Borchert2015N},  silver NW~\cite{Large2016SR,Khanarian2013JAP,Lee2008NL}, CNTs~\cite{Sapkota2017P,Abdellah2011IEEE,Forro2013AIPA,Majidian2017SR,Albert2012IEEE,Graf2016} and for hybrid magnetite-silver microparticles~\cite{Mietta2014JCP}. In most cases,  the distribution of filler lengths was well fitted by a lognormal distribution. Both the lognormal distribution and the Weibull distribution approximate well to the statistical data for the lengths of SWCNTs, nevertheless, the Weibull distribution is more reasonable~\cite{Wang2006}.
Khanarian et al.~\cite{Khanarian2013JAP} concluded that the ratio of the mean diameter to the weighted average of the lengths is important for the electrical properties of such films.
A bimodal distribution of CNT lengths has been created by mixing nanotubes taken from the long and short of batches of as-milled CNTs, and the electrical conductivity of the composite has been measured at different values of the weighted average of the lengths; the results being compared with those of the  as-milled samples~\cite{Majidian2017SR}. The as-milled samples showed electrical conductivity two orders of magnitude higher in the region of large values of the weighted averages of the lengths.

The electrical conductivity of a composite containing cylinders has been simulated~\cite{White2010AFM}
to enable the dependence of the electrical conductivity on the aspect ratio of the cylinders to be  studied. With the assumption that all the electrical resistance results from contact resistances between neighboring cylinders, calculations of the network conductivity for cylinders with finite dimensions were also performed. The authors compared the results of the simulation with experimental data, as well as with the conclusions of the analytical models (soft-core and core-shell) and found that the experimental  values agreed more closely with the simulations than with either of the analytical models.

Study of the transition to a resistance network has been carried out both with regard to the resistance of the sticks, and by taking into account the contact resistance, and the type of contact: metal--metal, metal--semiconductor, semiconductor--semiconductor~\cite{Behnam2007PRB}.
The network resistance has been calculated in the light of both the NW--NW junction resistance and the resistivity of the NW itself, while the model additionally considered the curviness of NWs~\cite{Hicks2018JAP}. For randomly oriented networks, curviness was found to be undesirable since it increased the resistivity. By contrast, for well-aligned networks,  some curviness is highly desirable, since, here, the resistivity minimum occurs for partially curvy NWs.

Two models with different distributions were investigated~\cite{Mutiso2012PRB}. In the first model, the rod lengths and diameters were almost normally distributed, i.e. negative values produced by the normal distribution were truncated. This led to breaking of the symmetry of the Gaussian distribution. In the second model, two fixed size populations of low- and high-aspect-ratio rods were considered. In calculating the resistance of the composite, only the contact resistance between the two rods was taken into account. The following result was obtained: the percolation threshold is insensitive to dispersity for small standard deviations (less than 0.3) in respect of length, but greatly decreases with increasing standard deviations (greater than 0.3). Compare this with the dispersed case, where, because of the above mentioned asymmetry of the length distribution,  there is an excess of longer  rods. Excellent agreement between the simulations and predictions from the generalized excluded-volume model solution for disperse rod networks has been observed. Systems of bimodal rod networks, indicate that the percolation threshold is lowered by increasing the relative volume fraction of the longer rods, and the longer these rods, the more pronounced the reduction at small relative volume fractions. A corresponding result was obtained for  a bimodal system of rods~\cite{Kyrylyuk2008PNAS}. The presence of a fraction of isotropically oriented rods of small aspect ratio has been shown to lower the percolation threshold for systems in which the longer rods are strongly aligned~\cite{Chatterjee2014JCP}.
A computer simulation study on the effects of length dispersity on geometrical percolation in suspensions of hard spherocylinders has been reported~\cite{Meyer2015JCP}. The results for bimodal, Gaussian and Weibull distributions have been compared and such comparisons have shown that the different kinds of dispersity exhibit non-trivial universal behavior.

A lognormal distribution of rod lengths was used in the simulations. It was  found that length dispersities of about 0.20 led to a roughly 20\%  reduction of the resistance in comparison to a system of NWs without dispersion and with the same  average length of NWs~\cite{Borchert2015N}.
In a model~\cite{Mietta2014JCP} where  random stick systems with lognormal length distribution were investigated, an increase in the value of the standard deviation led first to a slight decrease in the percolation threshold, then to a situation where, practically, no variation with the value of the standard deviation could be observed. The resistivity in percolation networks of one-dimensional elements with a lognormal length distribution~\cite{Hicks2009PRE}, have been studied using Monte Carlo (MC) simulations. For junction resistance-dominated random networks, the resistivity correlated with the root mean-square element length, whereas for element resistance-dominated random networks, the resistivity scaled with the average element length. The relationship between the critical density and the length of the nanotubes in the percolation paths has been obtained by combining MC and SPICE (Simulation Program with Integrated Circuit Emphasis)
simulations~\cite{Kang2011pssb}. Gaussian length distribution was used. In a mixed network of metallic and semiconducting nanotubes, the effects of the junctions determined the electrical properties of the network. The junctions were classified as Ohmic or Schottky contacts, depending on how the pairs of nanotubes formed. Each channel was modeled as a resistor, and each junction was modeled as two diodes connected in parallel with reverse polarity (diode ring) for a Schottky contact or as a resistor for an Ohmic contact.

Charge transport in a network cosisting only of semiconducting SWCNTs has been modeled as a
random resistor network (RRN) of tube--tube junctions~\cite{Schiessel2017PRM}. The impacts of both the average length and the standard deviation of the length on the network mobility were analyzed using a gamma distribution as the length distribution function.

In the above models, the resistance of the host matrix was assumed to be infinitely large. The effect of the dielectric properties of the host matrix on the percolation threshold when calculating the tunnel current has been investigated~\cite{Kyrylyuk2008PNAS}.

Recently, simulations of the electrical conductivity of two-dimensional (2D) systems with rod-like fillers of equal length have been performed both in lattice~\cite{Tarasevich2016PRE,Tarasevich2017PhysA,Tarasevich2018JPhCS} and in continuous~\cite{Tarasevich2018PREa,Tarasevich2018PREb} approaches.  In the present work, we have examined the effect of dispersity of filler length on the electrical conductivity of 2D composites with aligned rod-like fillers  by means of computer simulation.

The rest of the paper is constructed as follows. In Section~\ref{sec:methods}, the technical details of the simulations and calculations are described. Section~\ref{sec:results} presents our main findings. Section~\ref{sec:concl} summarizes the main results.

\section{Methods}\label{sec:methods}
\subsection{Preparation of the film samples}
We used a continuous approach to prepare each film sample. Zero-width (widthless) rod-like particles were deposited uniformly with given anisotropy onto a substrate  of size $L \times L$ and with periodic boundary conditions (PBCs). Intersections of the particles were allowed. The length of the particles, $l$, varied according to a log-normal distribution with the probability density function (PDF)
\begin{equation}\label{eq:lognorm}
  f(l)=\frac{1}{l\sigma_l\sqrt{2\pi }}\exp \left( -\frac{\left( \ln l-\mu_l \right)^2}{2\sigma_l^2} \right).
\end{equation}
The mean, $\langle l \rangle$, and the standard deviation, $\mathrm{SD} (l)$, are connected with the parameters of the log-normal distribution, $\mu_l$, $\sigma_l$,  as follows
\begin{equation}\label{eq:mean}
  \langle l \rangle = \exp\left(\mu_l+\frac{\sigma_l^2}{2}\right),
\end{equation}
\begin{equation}\label{eq:var}
  \mathrm{SD} (l)^2 = \left(\exp\left(\sigma_l^2\right)-1\right) \exp \left(2\mu_l+\sigma_l^2\right).
\end{equation}
A change of any parameter affects both the mean and the standard deviation. To avoid a superposition of different effects, the mean was set as a constant during simulation. In this case, we could extract and study the individual effect of the length dispersity. All our computations were performed for $
\langle l \rangle = 1.
$
For this particular value of the mean the parameters of the log-normal distribution are
$$
\mu_l = -\frac{\sigma_l^2}{2}, \quad  \sigma_l^2 = \ln \left(\mathrm{SD} (l)^2 + 1\right).
$$

The anisotropy of the system is characterized by  the order parameter
\begin{equation}\label{eq:s}
s=N^{-1}\sum_{i=1}^N \cos 2\theta_i,
\end{equation}
where $\theta_i$ is the angle between the axis of the $i$-th rod and the horizontal axis $x$, and $N$ is the total number of rods in the system (see, e.g.,~\cite{Frenkel1985PRA}). In our simulations, the angles were distributed according to a normal distribution\cite{Tarasevich2018PREa}
\begin{equation}\label{eq:angledistrib}
  f(\theta) = \frac{1}{\sqrt{- \pi \ln s}} \exp\left(\frac{\theta^2}{\ln s} \right).
\end{equation}

For each sample, a sequence of random positions (two coordinates for each rod), orientations, and lengths was generated. This sequence was used to produce a film with the desired number density of rods, $n$, i.e., the number of fillers per unit area,
\begin{equation}\label{eq:numdens}
  n = \frac{N}{L^2}.
\end{equation}

\begin{figure}[!htbp]
  \includegraphics[width=\columnwidth]{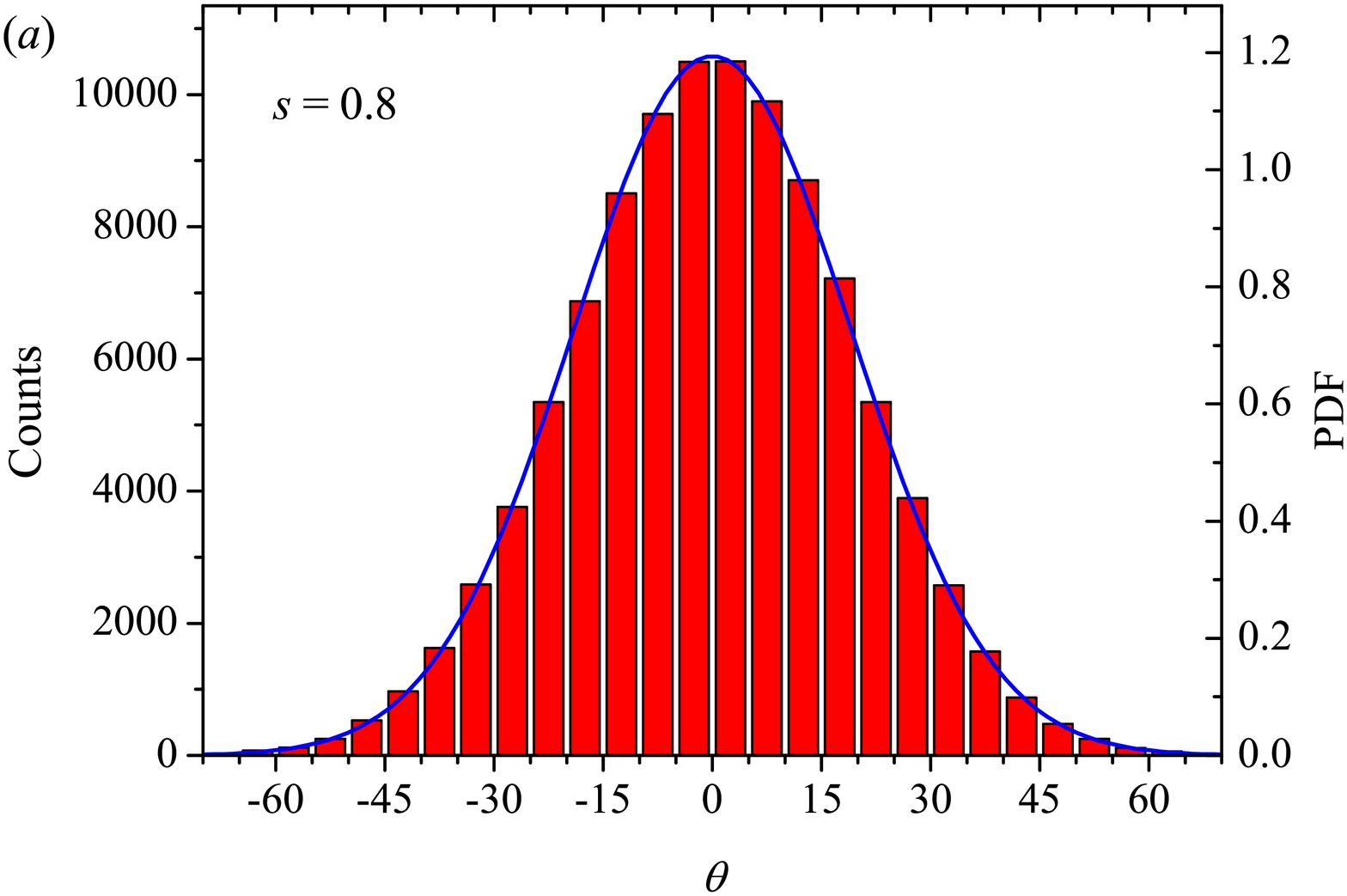}\\
  \includegraphics[width=\columnwidth]{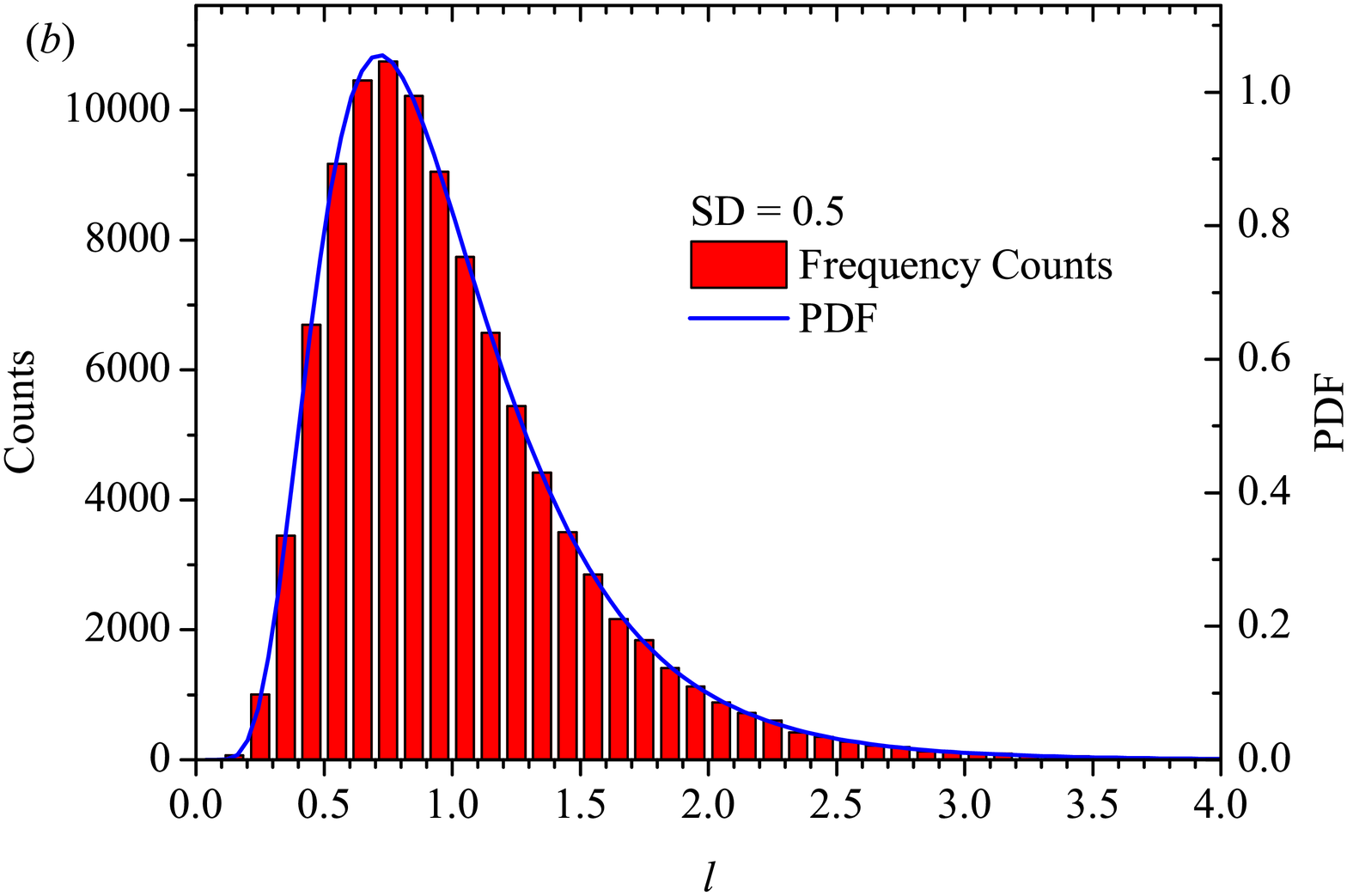}\\
  \caption{\label{fig:histograms}Examples of distributions of rods by angle and by length in one particular sample of a film with anisotropy $s=0.8$; the total number of rods is $N = 102400$. ($a$)~Angle distribution. The PDF of the normal distribution~\eqref{eq:angledistrib} is shown at an  appropriate scale for comparison. ($b$) Length distribution; the standard deviation $\mathrm{SD} = 0.5$. The PDF of the log-normal distribution~\eqref{eq:lognorm} is shown at an appropriate scale for comparison.}
\end{figure}

Figure~\ref{fig:histograms} demonstrates the angle- and the length-distribution of the rods in one particular sample of the film.

We performed our simulations for different values of the order parameter and length dispersity, viz., $s=0.5, 0.8, 1$, and $\mathrm{SD} = 0, 0.1, 0.5, 1.0$.

\subsection{Computation of the electrical conductivity}
The rods were considered to be highly conductive whilst the substrate was assumed to be poorly conductive. In all our computations, the electrical contrast, i.e., the ratio of the conductivity of the rods, $\sigma_p$, to the conductivity of the substrate, $\sigma_m$, was $\Delta = 10^6$. To account the electrical conductivity of both the substrate and the fillers, a three-step transformation of the samples was performed~\cite{Lebovka2018PRE,Tarasevich2018PREa,Tarasevich2018PREb}.

\paragraph{In the first step,} a sample was divided into square cells of equal size, in such a way that the film was transformed into a mesh of $m \times m$  cells.

\paragraph{In the second step,} a cell was marked as conductive when it contained any part of a rod or some parts of several rods, otherwise the cell was marked as insulating. In this way, the continuous problem of rods was transformed into a problem of linear polyominoes of typical length $\approx m/L$~\cite{Lebovka2018PRE,Tarasevich2018PREa,Tarasevich2018PREb}. The fraction of conductive cells was denoted as the concentration or the filling fraction, $p$. In the case of strictly aligned rods of equal size ($s=1$, $\mathrm{SD} = 0$), the polyominoes are simply rectangles with the aspect ratio~\cite{Tarasevich2018PREa}
\begin{equation}\label{eq:k}
k = m/L + 1.
\end{equation}

\paragraph{In the third step,} the PBCs were removed, i.e., the torus was unfolded into a plane. The Hoshen--Kopelman algorithm~\cite{HK76} was applied to check whether there were any percolating (spanning) clusters of the conductive cells.  Then, each cell was replaced by four conductors connected as a four-pointed star (crosswise). All four conductors were assumed to have the same conductivities, viz., $\sigma_p = 2 \cdot 10^6$ a.u. when the cell was marked as conductive and $\sigma_m = 2$ a.u. otherwise. In such a way the film was transformed into an RRN. The conductivity of such an RRN was calculated using the Frank--Lobb  algorithm~\cite{FL88} for each desired number density in both mutually perpendicular directions.  As a result, the electrical conductivity vs number density, $\sigma_i(n)$, the electrical conductivity vs  concentration, $\sigma_i(p)$, and the presence of the percolation cluster vs number density and concentration were obtained for each sample. Here, $i=\parallel, \perp$ means the directions along and perpendicular to the alignment of the fillers, respectively. For each pair  of values $s$ and $\mathrm{SD}$, the electrical conductivities were computed for the meshes $m = 256, 512, 1024$.

\subsection{Averaging of the electrical conductivity}\label{subsec:cond}
\begin{figure*}[!htbp]
  \centering
  \includegraphics[width=\columnwidth]{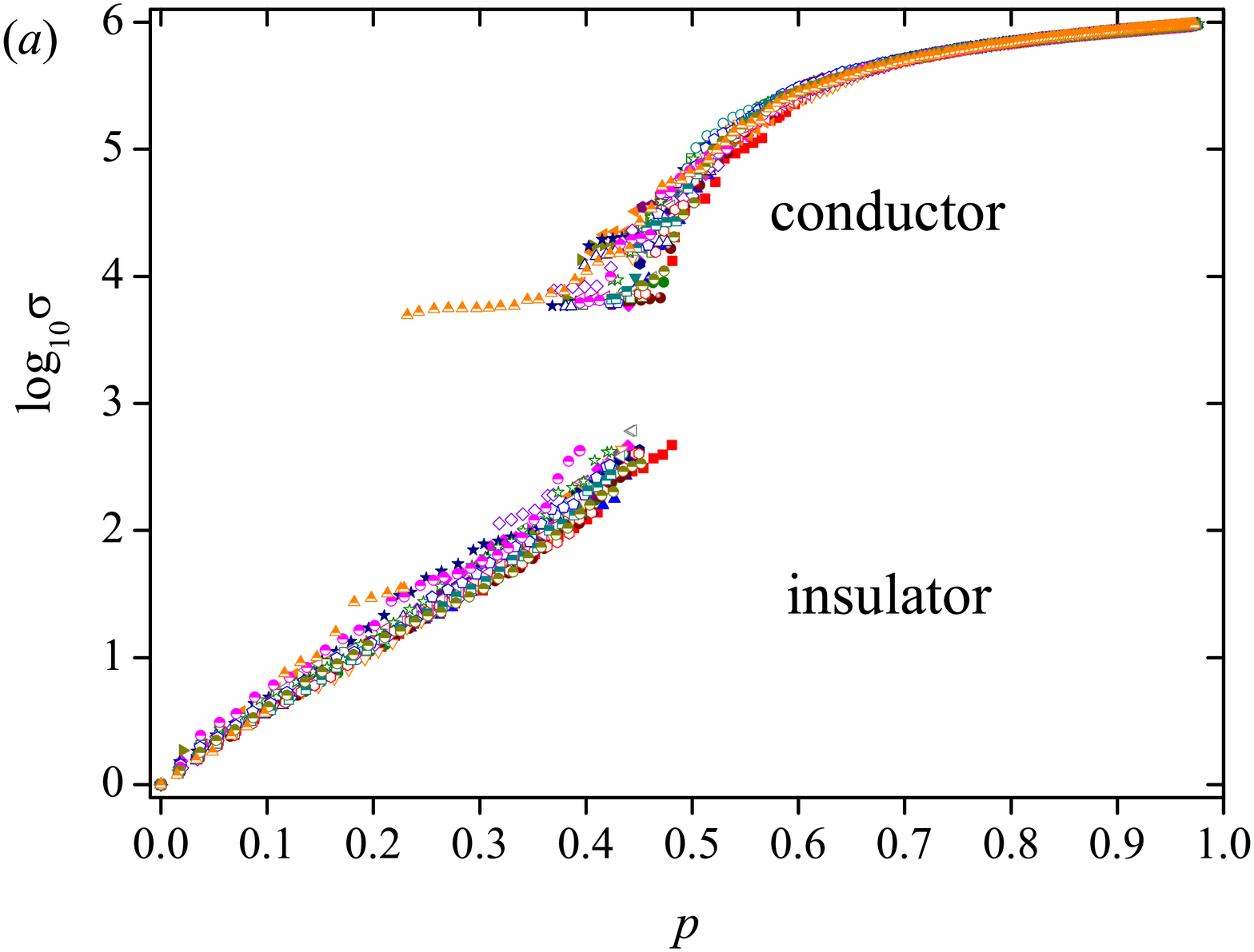}\hfill\includegraphics[width=\columnwidth]{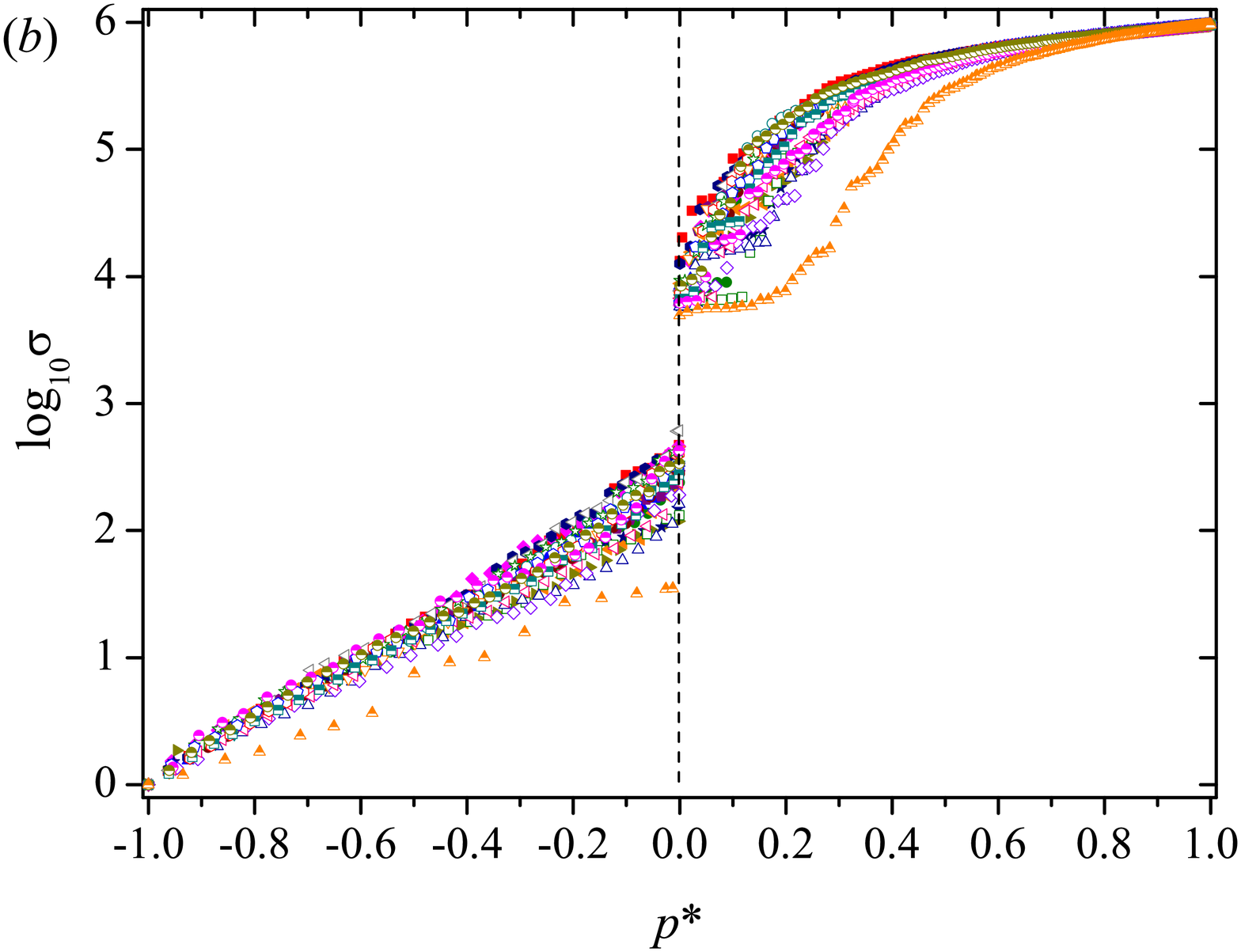}\\
  \includegraphics[width=\columnwidth]{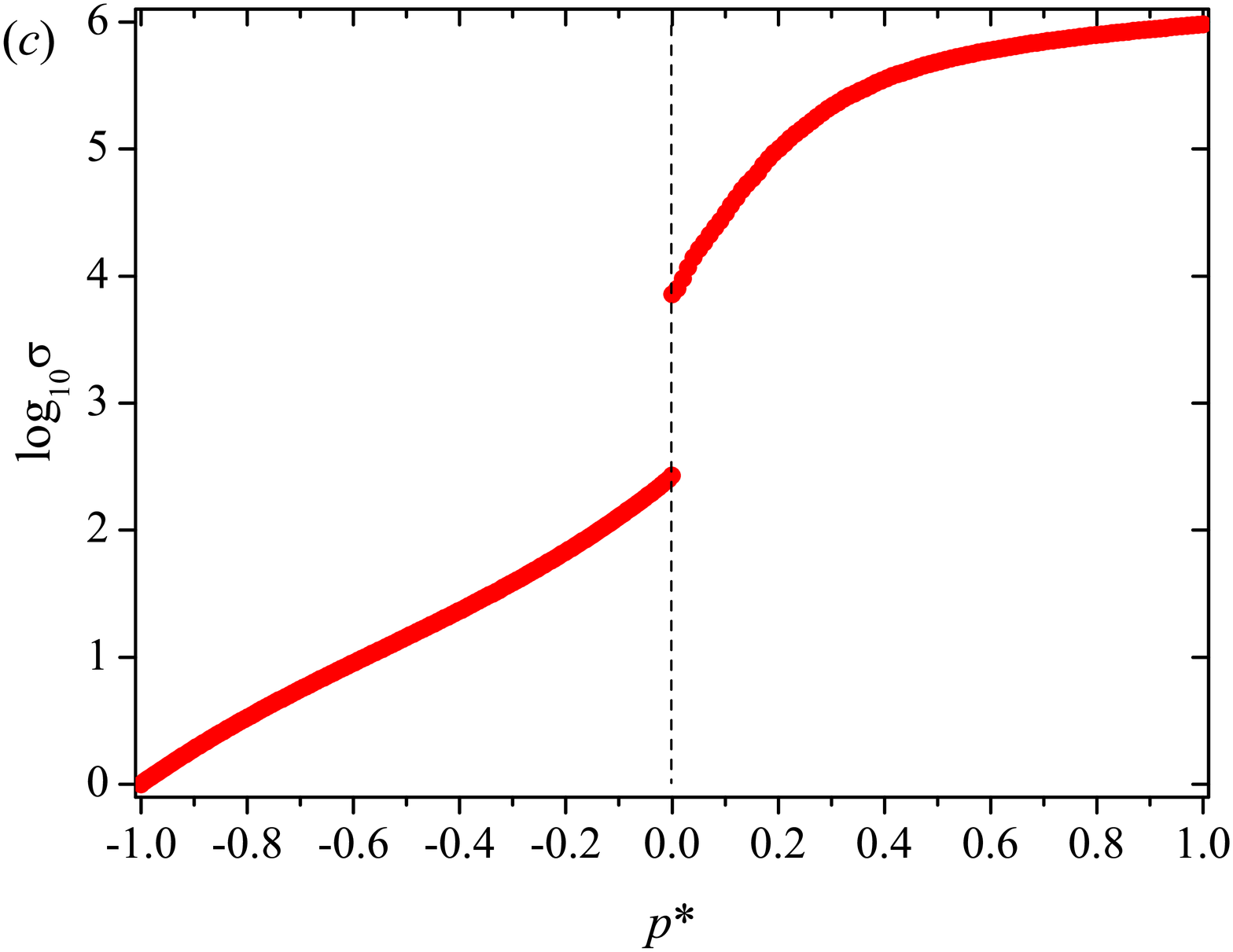}\hfill\includegraphics[width=\columnwidth]{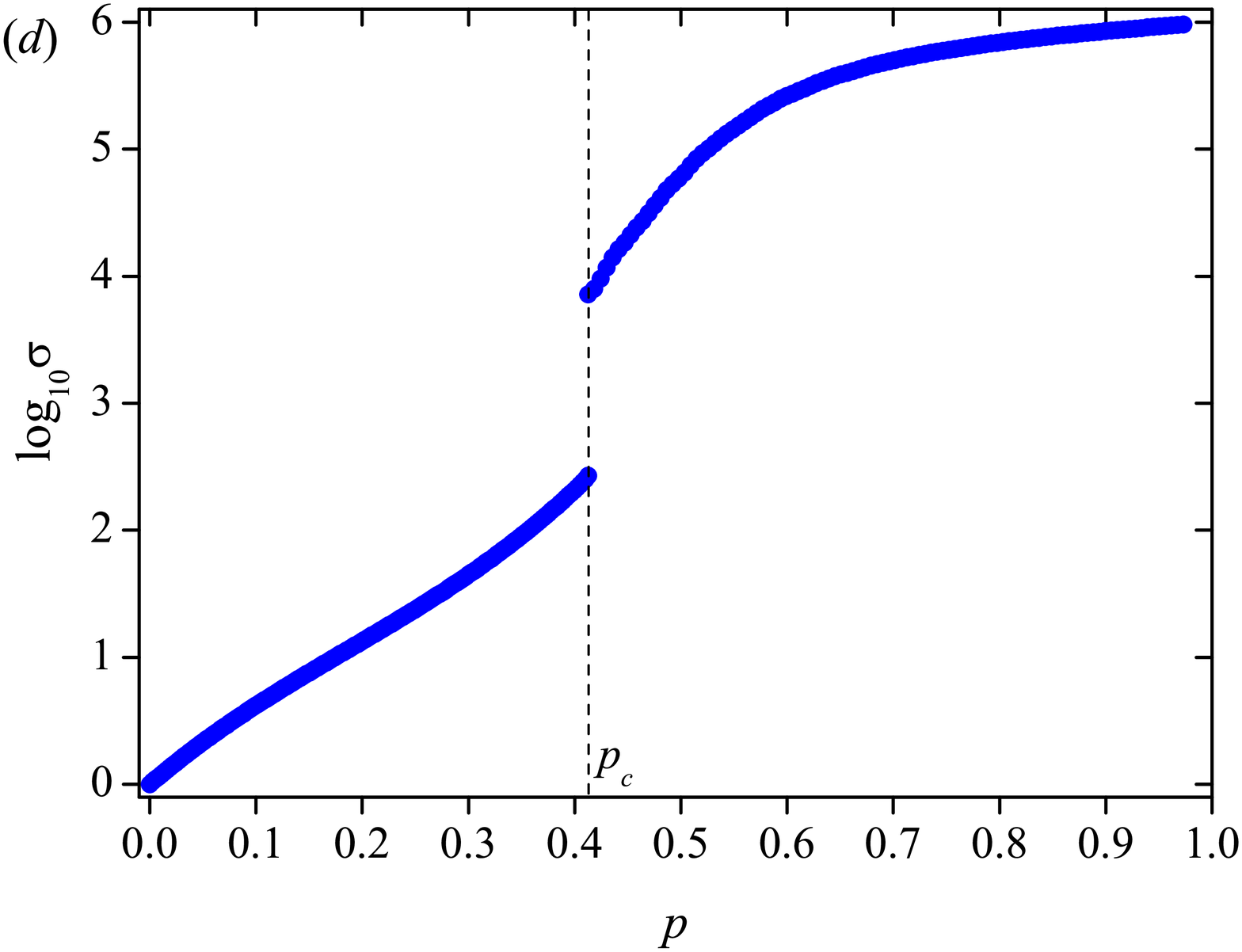}\\
  \caption{\label{fig:steps}Example of averaging of the electrical conductivity over independent runs.
  ($a$)~Step~1. All calculated values of the electrical conductivity, $\sigma_\parallel$, vs concentration, $p$, for 25  independent runs.
  ($b$)~Step~2.~Using Eq.~\eqref{eq:maps}, the results are mapped to the interval $[-1,1]$ where 0 corresponds to the percolation threshold, $p_c$.
  ($c$)~Step~3.~The averaged conductivity vs normalized concentration, $p^\ast$.
  ($d$)~Step~4.~The averaged conductivity vs concentration, $p$; the transformation $p^\ast \mapsto p$ has been performed using Eq.~\eqref{eq:recmaps}. $s=1$, $m=256$, $\mu_l = 0$, $\sigma_l = 0.75$. 
  }
\end{figure*}

The main concept, here, is that averaging the values of a physical quantity that characterizes different physical states, is a physical nonsense. Although further speculations are about the electrical conductivity, they can be applied to other quantities. Due to the random positions, orientations, and lengths of the fillers, at a concentration of fillers close to the percolation threshold, the electrical conductivity can correspond to an insulating state in one test and to a conducting state in another. Thus, to overcome this issue, our applied method of averaging of the electrical conductivity included four steps (Fig.~\ref{fig:steps}).

\paragraph{Step 1.} In the first step, the data for all the independent runs  were collected (Fig.~\ref{fig:steps}(a)). A curve of the electrical conductivity of any film vs concentration of fillers can be divided into two branches; one (lower) branch corresponds to the insulating states, whilst the other (upper) one corresponds to the conductive states (Fig.~\ref{fig:steps}(a)). These branches should be averaged separately over all the independent runs since the electrical conductivities below and above the phase transition point, in general, are described by different dependencies on the concentration of the conducting particles. A jump of the electrical conductivity at the percolation threshold arises due to the finite size of the system under consideration; this jump decreases as the size of the system increases and vanishes in the thermodynamic limit.

\paragraph{Step 2.}
Our method is based on the change of variables similar to that described in~\cite{Snarskii2016}.
Note that the transformation
\begin{equation}\label{eq:p1}
  p \to \frac{p-p_c}{p_c }
\end{equation}
maps the interval $[0,p_c]$ into the interval $[-1,0]$. Similarly, the transformation
\begin{equation}\label{eq:p2}
  p \to \frac{p-p_c}{1 - p_c}
\end{equation}
maps the interval $[p_c,1]$ into the interval $[0,1]$. Although the percolation threshold $p_c$ has a  definite value in the thermodynamic limit, any simulations are always carried out on finite systems, and this leads to a dispersion of the value of this quantity from one experiment to another. The percolation threshold should by precisely defined to the resolution of a single added rod for each particular deposition sequence of rods. For each sample, the curve $\sigma(p)$ was divided into two branches at the percolation threshold. The lower branch was mapped to the interval $[-1,0]$ using  transformation~\eqref{eq:p1} while the upper branch was mapped to the interval $[0,1]$ using  transformation~\eqref{eq:p2} (Fig.~\ref{fig:steps}(b)). Thus,
\begin{equation}\label{eq:maps}
  p^\ast =
  \begin{cases}
    \dfrac{p-p_c^{(i)}}{p_c^{(i)} }, & \mbox{if } p <  p_c^{(i)}, \\
    \dfrac{p-p_c^{(i)}}{1 - p_c^{(i)}}, & \mbox{if } p \geq  p_c^{(i)},
  \end{cases}
\end{equation}
where $p_c^{(i)}$ is the percolation threshold for the $i$-th sample.

\paragraph{Step 3.}
Each branch was averaged over all the independent runs (Fig.~\ref{fig:steps}(c)). Linear interpolation between the experimental points was used for this with the averaging being carried out for the desired values of concentrations.

\paragraph{Step 4.}
After averaging, both branches were transferred to the original interval of concentrations with abscises of the transformations being reciprocal to the transformations~\eqref{eq:p1}  and \eqref{eq:p2} (Fig.~\ref{fig:steps}(d)), viz.,
\begin{equation}\label{eq:recmaps}
  p =
  \begin{cases}
    \langle p_c \rangle + \langle p_c \rangle p^\ast, & \mbox{if } p^\ast <  0, \\
    \langle p_c \rangle + p^\ast - p^\ast \langle p_c \rangle, & \mbox{if } p^\ast \geq  0,
  \end{cases}
\end{equation}
where $\langle p_c \rangle = N^{-1}_{ir} \sum_i p_c^{(i)}$ is the mean value of the percolation threshold averaged over all the independent runs and $N_{ir}$ is the number of independent runs.

In our computations, typically, the final curves were the result of averaging over 25 independent runs.
Throughout the text, the error bars in the figures correspond to the standard deviation of the mean. When not shown explicitly, they are of the order of the marker size.

\subsection{Intrinsic electrical conductivity}
For randomly oriented and arbitrarily shaped particles with electrical conductivity $\sigma_p$ suspended in a continuous medium with electrical conductivity $\sigma_m$, the following virial expansion~\cite{Douglas1995AdvChPh,Garboczi1996PRE} is valid
\begin{equation}\label{eq:Garboczi1996}
\frac{\sigma}{\sigma_m} =1+[\sigma]p+ \mathrm{O}(p^2),
\end{equation}
where $p$ is the volume fraction of the particles.
The quantity
\begin{equation}\label{eq:ic}
[\sigma] = \left.\frac{\mathrm{d}\ln\sigma}{\mathrm{d}p}\right|_{p \to 0},
\end{equation}
is called the ``intrinsic conductivity''.
The ``intrinsic conductivity'', $[\sigma]$, depends on different factors, such as the electrical conductivity contrast, the particles shape, their orientation with respect to the direction of measurement of the electrical conductivity, the spatial dimension, and the continuous or discrete nature of the problem.

\subsection{Anisotropy of the electrical conductivity}

Due to anisotropic deposition of the rods according to the given value of the order parameter $s$, the electrical conductivity of the film may be different in two mutually perpendicular directions. To characterize such an electrical anisotropy, we used the quantity~\cite{Tarasevich2016PRE,Tarasevich2017PhysA}
\begin{equation}\label{eq:delta}
  \delta = \frac{\log_{10} \left(\sigma_\parallel / \sigma_\perp\right)}{\log_{10} \Delta}.
\end{equation}

Note that, at the percolation threshold, the effective electrical conductivity of a two-phase thin film at equal concentrations of the phases and with a random distribution of them is equal to the geometric mean of the conductivity of the phases
\begin{equation}\label{eq:dykhne}
\sigma_g = \sqrt{\sigma_m \sigma_p},
\end{equation}
where  $\sigma_p$ is the electrical conductivity of the particles and $\sigma_m$ is the electrical conductivity of the host matrix (medium)~\cite{Dykhne1971,Efros1976pssb}. By contrast, for anisotropic systems, there is a significant difference between the electrical conductivities in the two mutually perpendicular directions. For a material with electrically isotropic particles of equal size and anisotropic shape aligned along one direction~\cite{Carmona1987PRB},
\begin{equation}\label{eq:jump}
\begin{aligned}
  \sigma_\parallel &= k^{2u} \sigma_p^{1 - u} \sigma_m^u,\\
  \sigma_\perp &= k^{2u - 2} \sigma_p^{1 - u} \sigma_m^u.
\end{aligned}
\end{equation}
Here, $k=l/d$ is the aspect ratio of the particles. $u = 1/2$ in two dimensions~\cite{Efros1976pssb}, hence, \eqref{eq:jump} reduces to~\cite{Balagurov2008}
\begin{equation}\label{eq:jump2d}
\begin{aligned}
  \sigma_\parallel &= k \sigma_g,\\
  \sigma_\perp &= k^{-1} \sigma_g.
\end{aligned}
\end{equation}

Hence, the ratio of the electrical conductivities at the percolation threshold is
\begin{equation}\label{eq:anisotropy}
  \frac{\sigma_\parallel}{\sigma_\perp} = k^2
\end{equation}
independent  of the electrical contrast $\Delta$~\cite{Shklovskii1978PSS,Troadec1981JPC,Carmona1987PRB}.

\section{Results}\label{sec:results}

\subsection{Finite size effect}
In finite size discrete systems with aligned rods of equal size, percolation in different directions occurs at different concentrations of the rods~\cite{Ackermann2016SR}.
Although the size of the substrate, $L$, has a small effect on the behavior of the electrical conductivity~\cite{Lebovka2018PRE}, it shifts the percolation threshold especially in the direction of rod alignment~\cite{Tarasevich2012PRE}.

\begin{figure}[!htbp]
  \centering
  \includegraphics[width=\columnwidth]{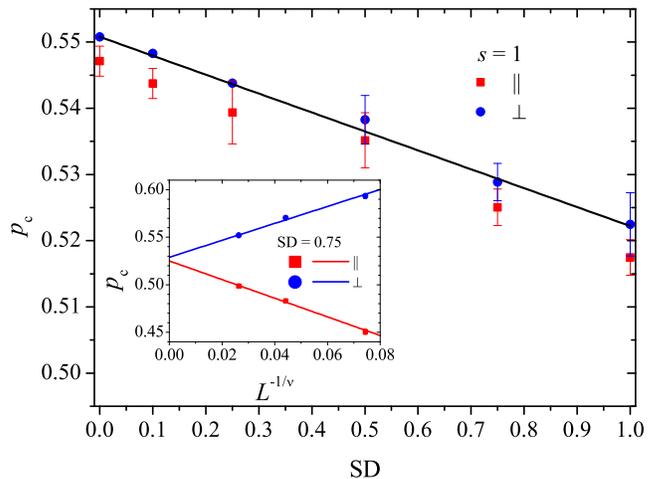}
  \caption{Percolation threshold, $p_c$, for completely aligned rods ($s=1$) vs the standard deviation, $\mathrm{SD}$, in the thermodynamic limit. The line corresponds to a linear fit. Inset: example of scaling of the percolation thresholds for the standard deviation $\mathrm{SD}=0.75$ and $m/L =8$. Here $\nu = 4/3$ is the critical exponent~\cite{Stauffer}.\label{fig:percscaling}}
\end{figure}

In the systems under considerations, the finite size effect is expected to be more pronounced  for strongly anisotropic systems, i.e.  for systems with completely aligned rods ($s=1$) (Fig.~\ref{fig:percscaling}).
The percolation thresholds along and perpendicular to the direction of the rod alignment are significantly different (see inset in Fig.~\ref{fig:percscaling}). When the size of the system increases, this difference between the values of the percolation thresholds decreases. This has an effect on the anisotropy of the electrical conductivity in the vicinity of the percolation threshold. In the thermodynamic limit, the percolation thresholds decrease linearly as dispersity increases. The dependence of the percolation threshold on the dispersity is well fitted by the linear function
\begin{equation}\label{eq:fit}
  p_c = p_{c0} - \alpha \cdot \mathrm{SD},
\end{equation}
where $p_{c0} =0.55079\pm 0.00006$ and $\alpha = 0.0286 \pm 0.0009$.

\begin{figure}[!htbp]
 \includegraphics[width=\columnwidth]{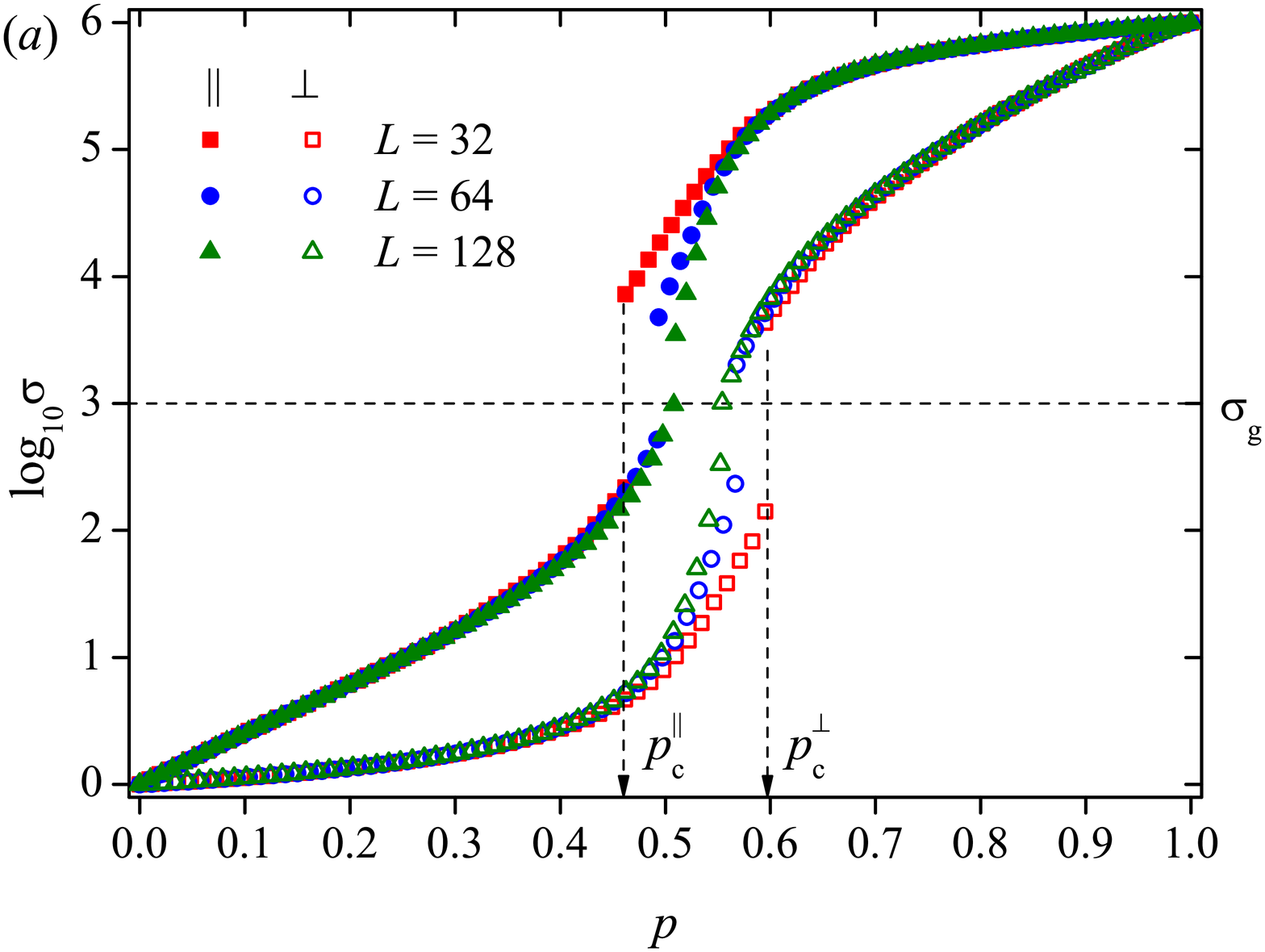}\\
 \includegraphics[width=\columnwidth]{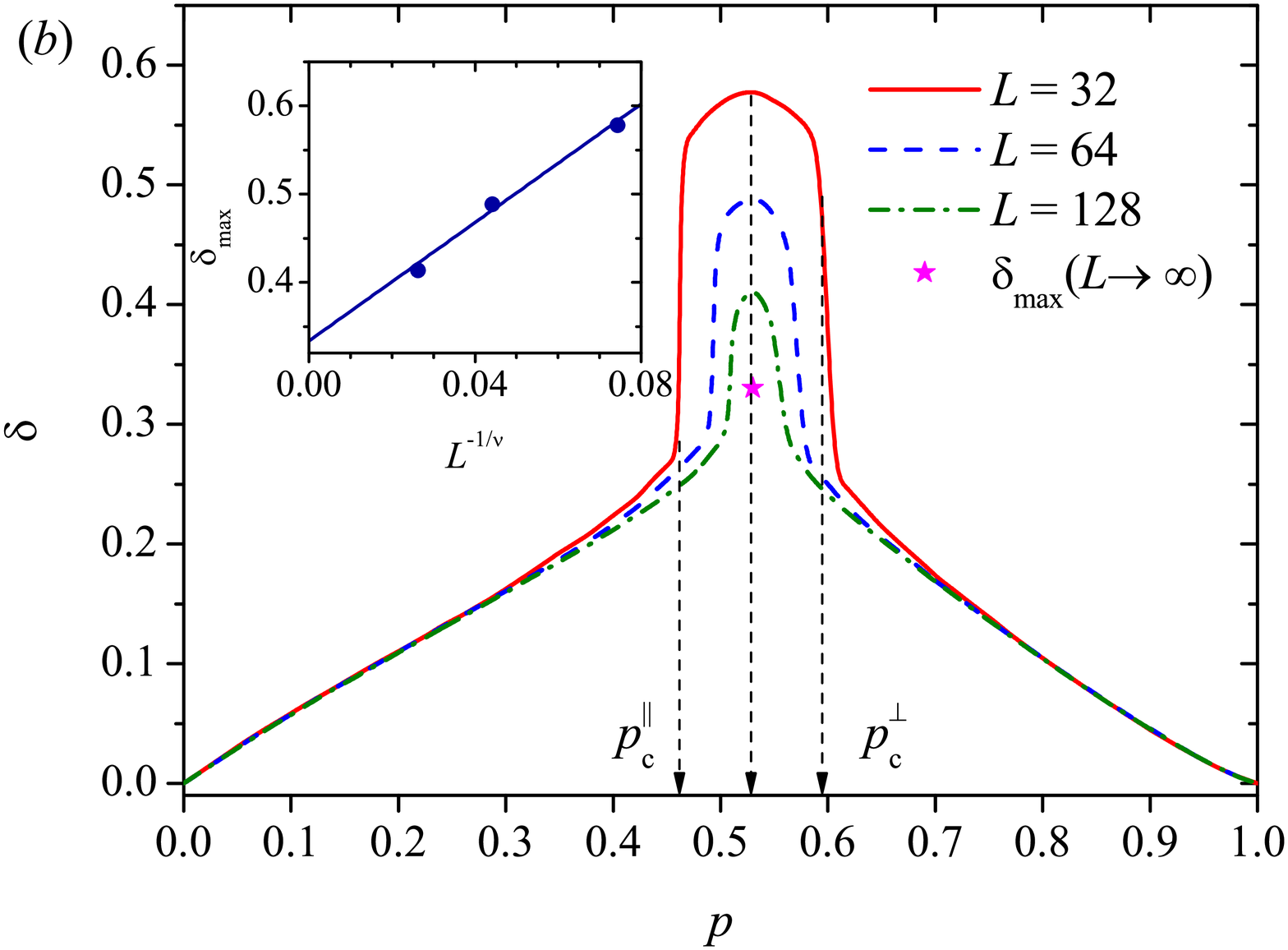}\\
 \caption{\label{fig:scaling}Examples of finite size effect for different values of $L$ and fixed ratio $m/L = 8$. Completely aligned rods ($s=1$) with length distribution corresponding to the log-normal distribution with the standard deviation $\mathrm{SD} = 0.5$.
 ($a$)~Averaged electrical conductivity, $\sigma$, vs the concentration, $p$. The arrows point out the percolation thresholds $p_c^\perp$ and $p_c^\parallel$ in two perpendicular directions for one particular system size $L=32$.
 ($b$)~Electrical anisotropy, $\delta$, vs the concentration, $p$. The additional arrow indicates the position of the maximal anisotropy, $\delta_\text{max}$. The star corresponds to the maximal anisotropy, $\delta_\text{max}$, in the thermodynamic limit $L \to \infty$. Inset: scaling of maximal anisotropy, $\delta_\text{max}$.}%
 \end{figure}

Figure~\ref{fig:scaling} demonstrates an example of the finite size effect on the electrical conductivity and its anisotropy for one particular value of the standard deviation $\mathrm{SD} = 0.5$. Figure~\ref{fig:scaling}($a$) shows that electrical conductivity is sensitive to the size of the system only in the vicinity of the percolation threshold. Although the effect is small, it produces a ``hump'' in the curve of the electrical conductivity (Fig.~\ref{fig:scaling}($b$)). The ``hump'' decreases as the size of the system increases. The inset in Fig.~\ref{fig:scaling}($b$)  illustrates how the maximal value of the electrical anisotropy, $\delta_\text{max}$, decreases with the system size, $L$. Note that the theoretical prediction for those systems with a standard deviation $\mathrm{SD}=0$ Eq.~\eqref{eq:anisotropy} gives the electrical anisotropy Eq.~\eqref{eq:delta} $\delta  = 0.301$ when $k=9$, i.e., $m/L=8$. This value is reasonably close to the $\delta_\text{max} = 0.330 \pm	0.017$ obtained for the system with moderate dispersity.

\subsection{Electrical conductivity}

All results presented in this section were obtained for systems of size $L=32$.

The dependencies of the intrinsic conductivity, $[\sigma]$, on the aspect ratio, $k$, are presented in Table~\ref{tab:IC} for films with completely aligned rods ($s=1$) and different values of the length dispersity. The transversal intrinsic conductivity is insensitive to the dispersity of rod length. In contrast, the longitudinal intrinsic conductivity increases as the standard deviation, $\mathrm{SD}$, increases. This behavior corresponds qualitatively to the estimations, which can be drawn from the theoretical results for systems with elongated particles of equal size and $\Delta \gg 1$~\cite{Balagurov2008},
\begin{equation}\label{eq:IClimit}
  \begin{aligned}
   \protect[\sigma_{\parallel}] &= k + 1, \\
     [\sigma_\perp] &= 1 + k^{-1},
   \end{aligned}
\end{equation}
where $k$ is the aspect ratio of the rods calculated according Eq.~\eqref{eq:k}.
\begin{table}[htb]
  \caption{\label{tab:IC} Intrinsic conductivities for different values of $m$ and $\mathrm{SD}$.}
\begin{ruledtabular}
    \begin{tabular}{cdddd}
\diagbox{$m$}{SD} & 0.0 & 0.1 & 0.5 & 1.0\\
    \hline
    \multicolumn{5}{c}{longitudinal}\\
    \hline
256 & 8.289	& 8.507 & 9.712 & 14.058\\
512 & 14.637	& 14.916 & 17.340 & 26.620\\
1024 & 26.778 & 26.865	& 31.839 & 50.011\\
    \hline
        \multicolumn{5}{c}{transversal}\\
    \hline
256 &  1.138 & 1.141 & 1.137 & 1.132\\
512  & 1.089  & 1.090  & 1.088  & 1.084\\
1024 & 1.055 & 1.058 & 1.057 & 1.055\\
  \end{tabular}
\end{ruledtabular}
\end{table}

\begin{figure*}[!htbp]
  \includegraphics[width=\columnwidth]{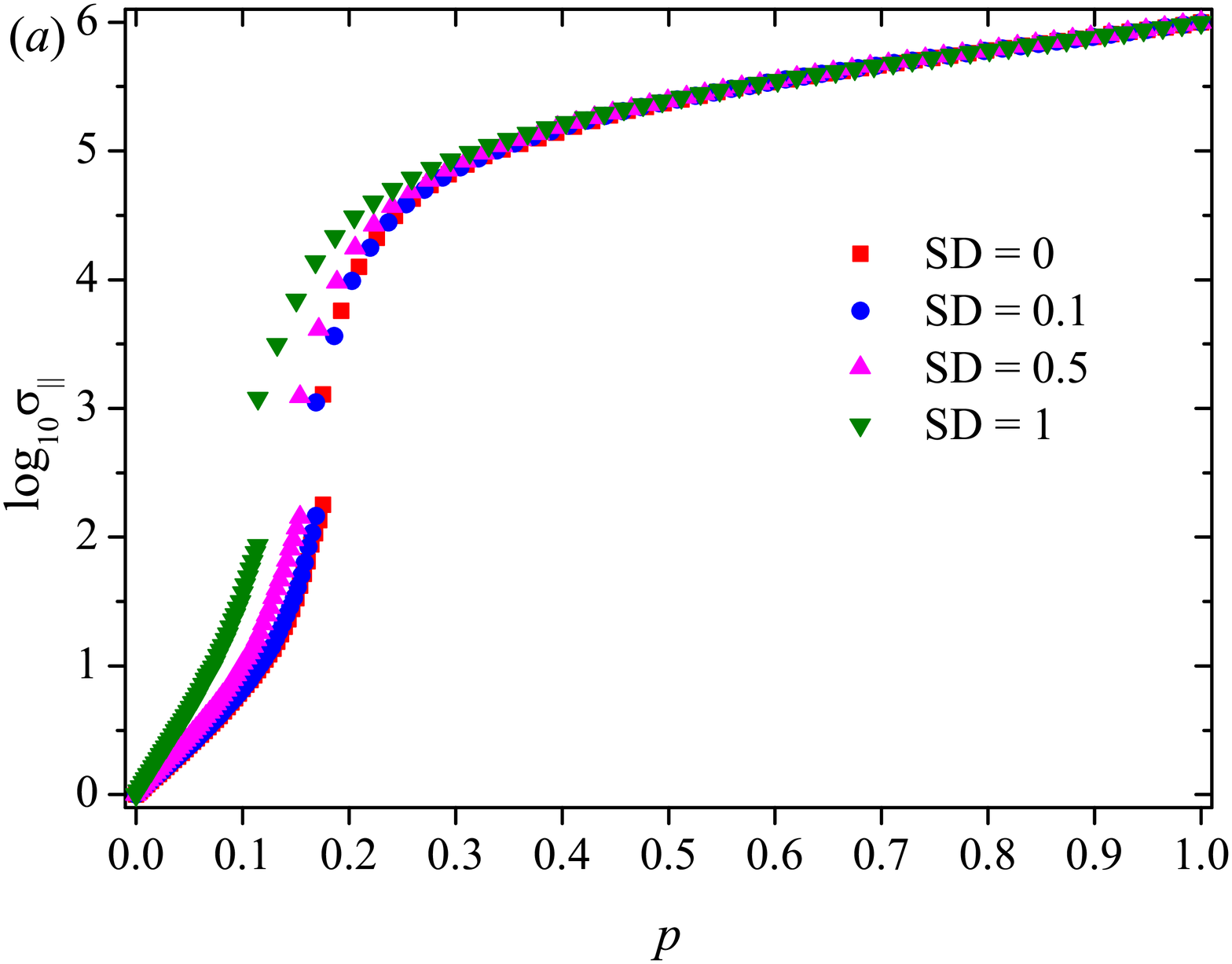}%
  \includegraphics[width=\columnwidth]{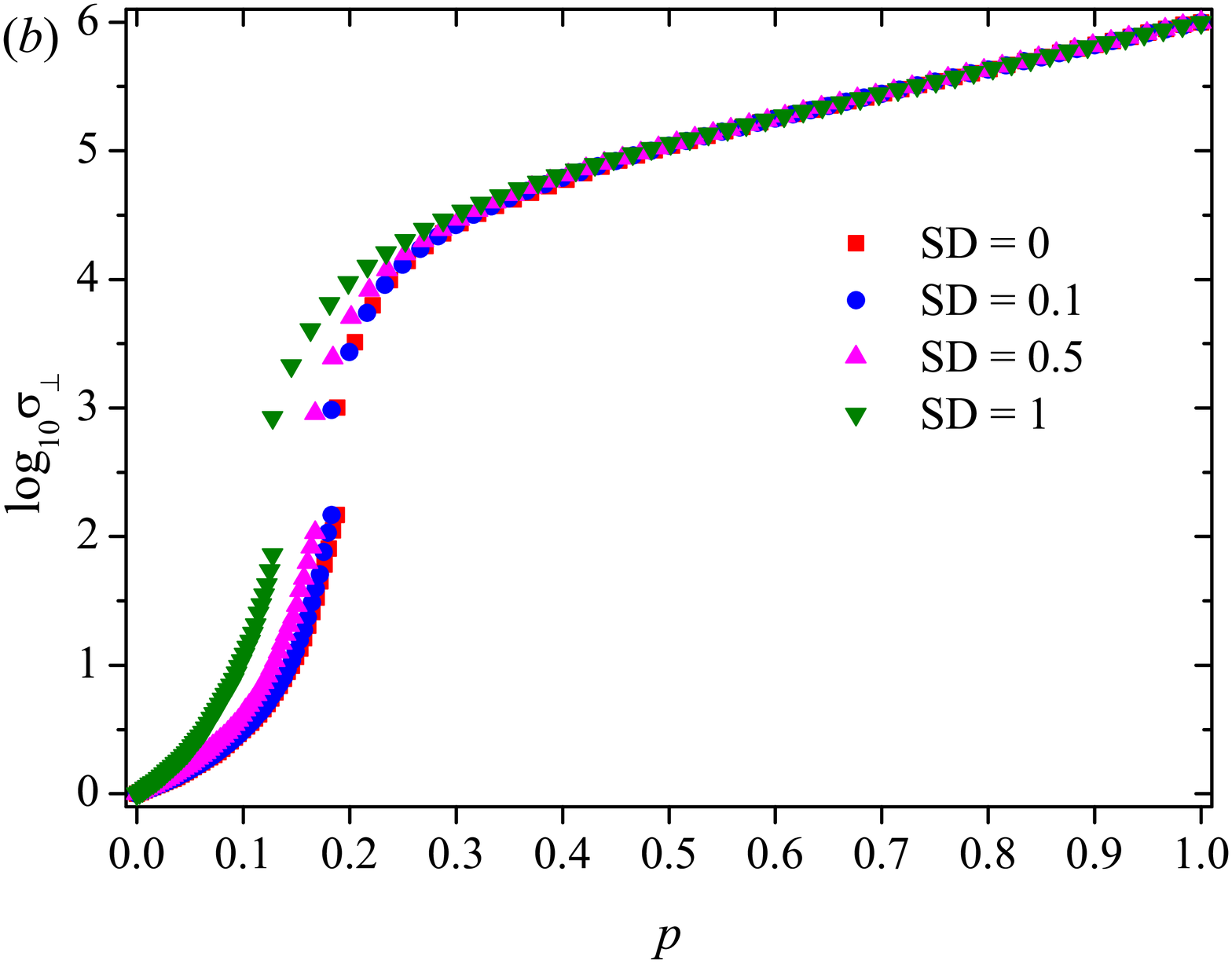}\\
  \caption{\label{fig:condSD} Examples of the dependencies of the effective electrical conductivity, $\sigma$, on the concentration, $p$, at $\mathrm{SD}=0, 0.1, 0.5, 1.0$ and $s=0.5$, $m=1024$, $L=32$: ($a$) along the direction of alignment of the rods; ($b$) in the perpendicular direction.}
\end{figure*}

Figure~\ref{fig:condSD} shows an example of the dependence of the effective electrical conductivity, $\sigma$, on the concentration, $p$,  along the direction of alignment of the rods (Fig.~\ref{fig:condSD}($a$)) and in the perpendicular direction (Fig.~\ref{fig:condSD}($b$)) for  $\mathrm{SD}=0, 0.1, 0.5, 1.0$, $m=1024$  and an intermediate value of the order parameter, $s=0.5$. The value of the standard deviation, $\mathrm{SD}$, affected the percolation thresholds in both directions. However, the effects were different in the parallel and transversal directions. In the case, where all the rods are aligned along one direction, $s=1$, the transversal effective electrical conductivity should not change with the variation in the dispersity of the length, and this is confirmed by the numerical results.

\begin{figure}[!htbp]
  \centering
  \includegraphics[width=\columnwidth]{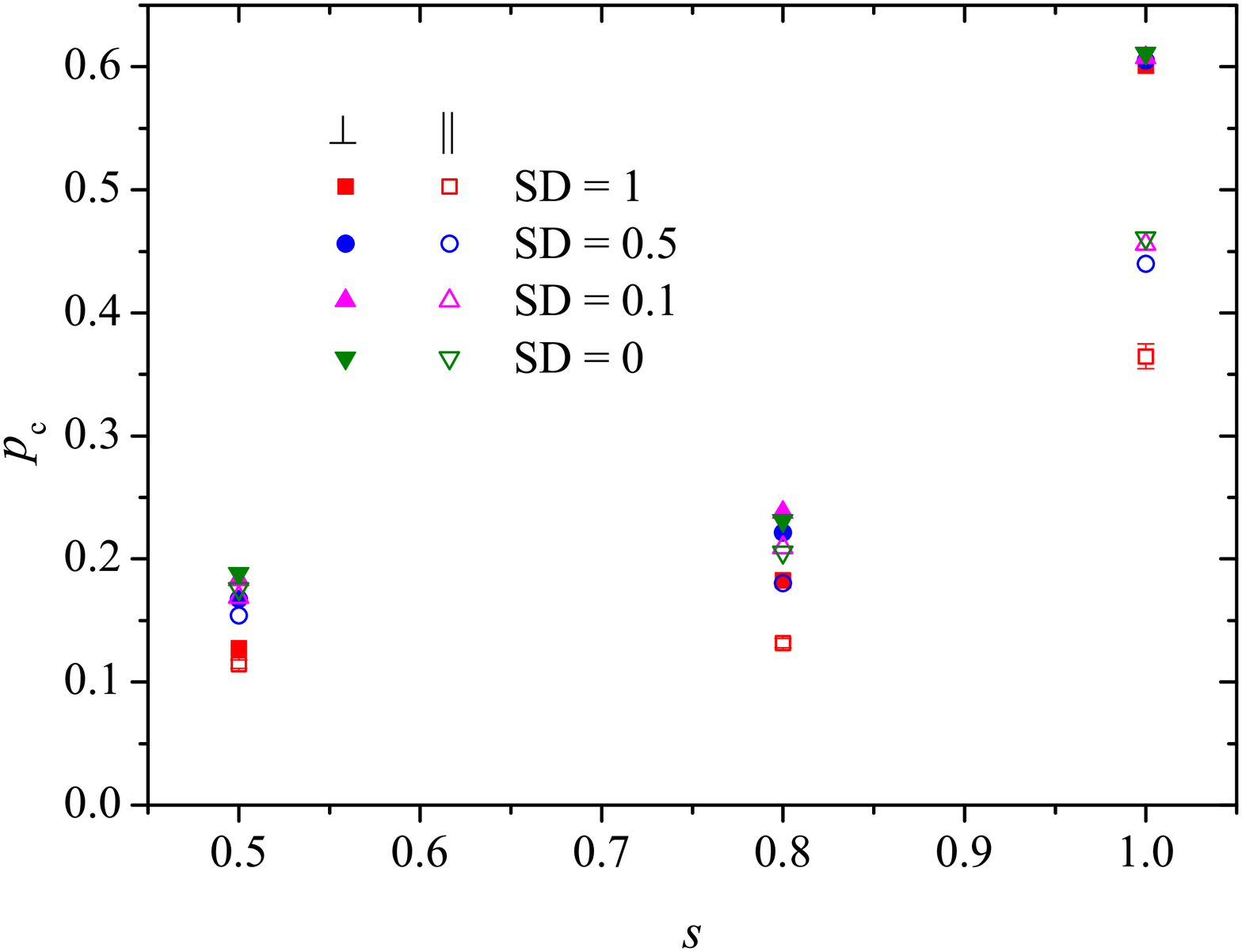}
    \caption{\label{fig:perc} Examples of the percolation threshold, $p_c$, versus the order parameter, $s$, for different values of the standard deviation $\mathrm{SD}$. $m=1024$, $L=32$.}
\end{figure}

Figure~\ref{fig:perc} shows the dependence of the percolation threshold, $p_c$, along the direction of alignment of the rods (closed symbols) and in the perpendicular direction (open symbols) on the order parameter, $s$, for four values of the standard deviation when $m=1024$. There is a general tendency for the percolation threshold value to increase when the order parameter increases.

\begin{figure}[!htbp]
  \includegraphics[width=\columnwidth]{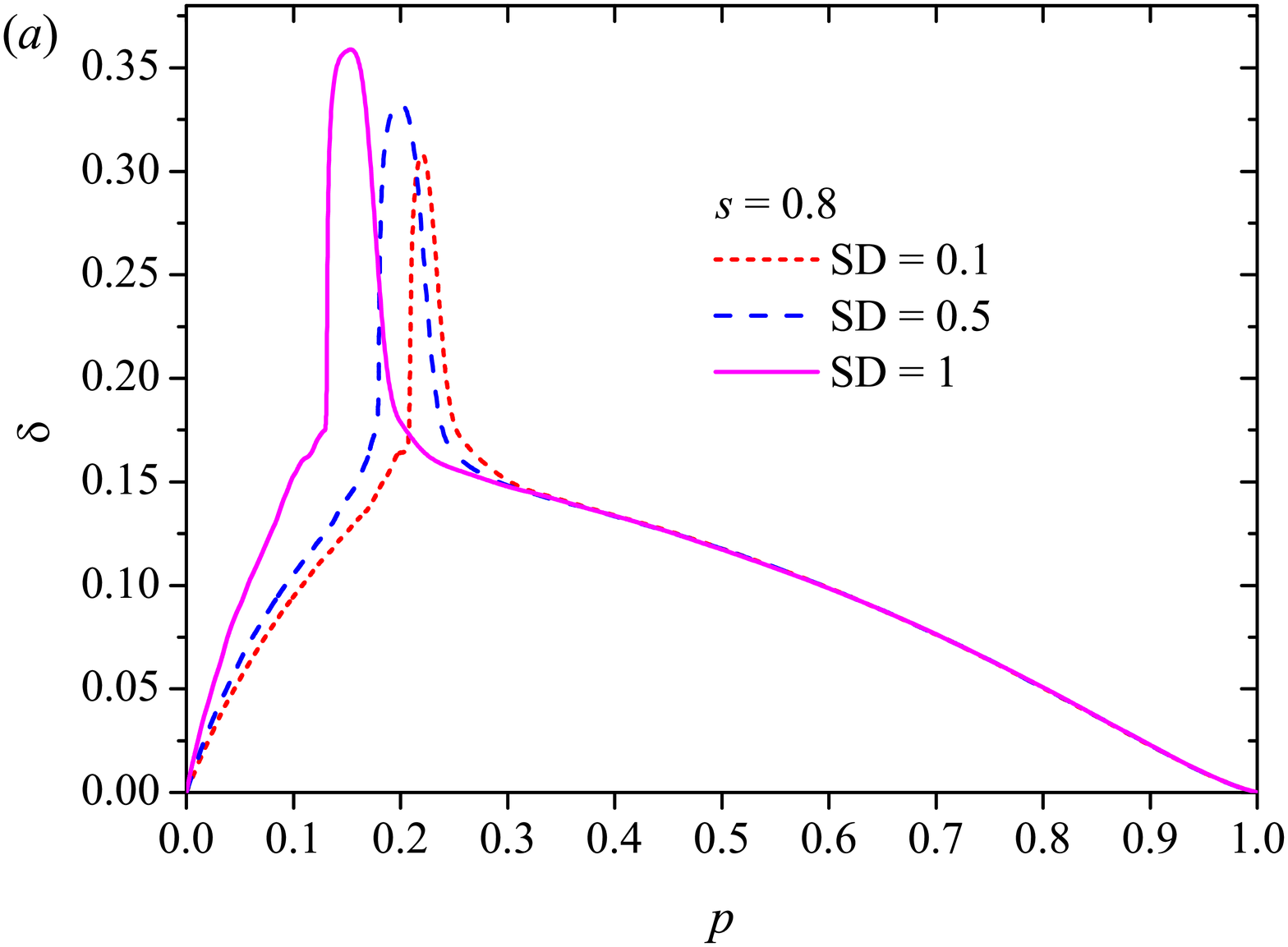}\\
  \includegraphics[width=\columnwidth]{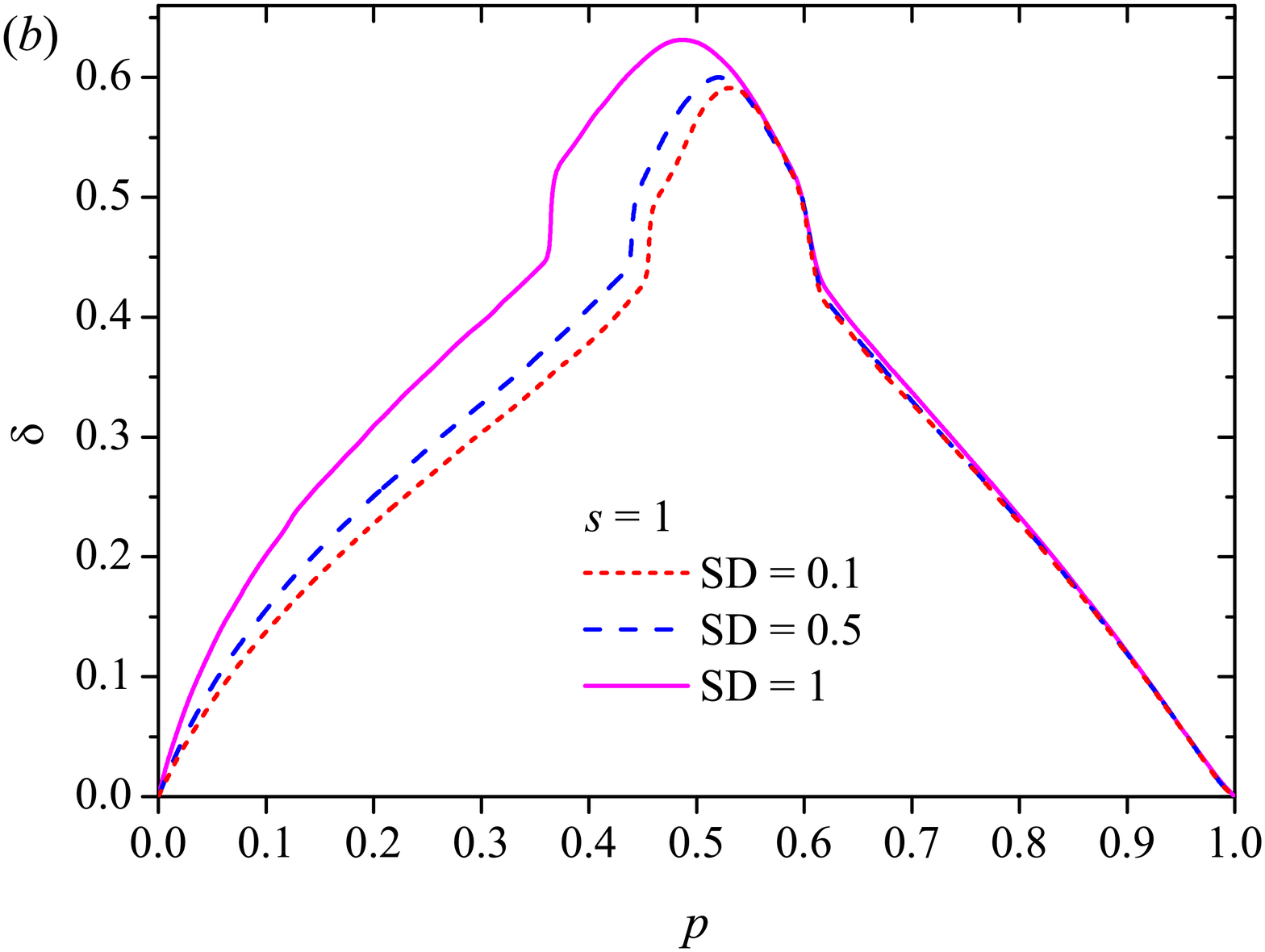}\\
  \caption{\label{fig:delta} Examples of the electrical anisotropy ratio, $\delta$, versus the concentration, $p$, for $\mathrm{SD}=0.1, 0.5, 1.0$, $m=1024$, $L=32$: ($a$) $s=0.8$; ($b$) $s=1$.}
\end{figure}

Figure~\ref{fig:delta} presents examples of the electrical anisotropy ratio, $\delta$,~\eqref{eq:delta} versus the concentration, $p$, for $\mathrm{SD}=0.1, 0.5, 1.0$, $m=1024$, and $s=0.8, 1$. In the cases of both partially aligned rods, $s<1$, (Fig.~\ref{fig:delta}($a$)) and completely aligned rods, $s=1$, (Fig.~\ref{fig:delta}($b$)), the maximum of the electrical anisotropy ratio became more pronounced and shifted to correspond to the smaller values of $p$ as the dispersity increased. The large the order parameter, $s$, the larger the electrical anisotropy, $\delta$.

\begin{figure}[!htbp]
  \centering
  \includegraphics[width=\columnwidth]{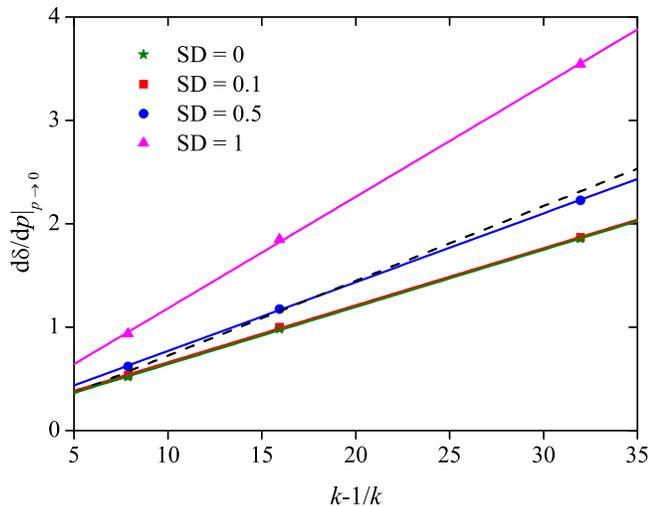}
  \caption{Growth rate of the anisotropy of the electrical conductivity vs $k-k^{-1}$ at small values of the concentration ($p \to 0$) for different values of the standard deviation, $\mathrm{SD}$, and $s=1$. Solid lines correspond to linear fits; the dashed line corresponds to Eq.~\eqref{eq:ddeltadp}.}\label{fig:ddeltadp}
\end{figure}

At small concentrations ($p \to 0$), the rate of growth of the anisotropy of the electrical conductivity increased as the length dispersity, $\mathrm{SD}$, and the aspect ratio, $k$, increased for films with completely aligned rods ($s=1$) (Fig.~\ref{fig:ddeltadp}). This behavior corresponds to the theoretical estimation
\begin{equation}\label{eq:ddeltadp}
  \left.\frac{\mathrm{d} \delta }{\mathrm{d} p}\right|_{p \to 0} = \frac{k - k^{-1}}{\ln \Delta}
\end{equation}
obtained using Eqs.~\eqref{eq:Garboczi1996}, \eqref{eq:delta}, and \eqref{eq:IClimit}. The slope increases as the standard deviation, $\mathrm{SD}$, increases. The difference between the theoretical prediction of the slope $\ln^{-1}\Delta$ probably arises due to the different shapes of the particles in our model (rectangles) and in the theory (ellipsoids).

It is noticeable that the length dispersity demonstrated no significant effect on the electrical conductivities in either direction when the concentration of fillers was above the value $p_\perp$ (see Fig.~\ref{fig:condSD} and Fig.~\ref{fig:delta}). This suggests a universality of the electrical conductivity, not only for dense isotropic RRNs~\cite{Kumar2016JAP} but also for dense anisotropic RRNs.

\begin{figure}[!htbp]
  \centering
  \includegraphics[width=\columnwidth]{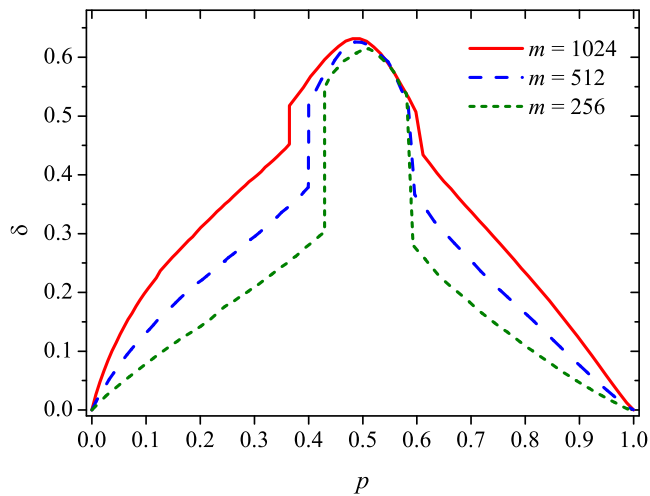}
  \caption{Example of the dependencies of the electrical anisotropy, $\delta$, vs the concentration, $p$, for different values of $m=256,512,1024$. Completely aligned rods ($s=1$) with high length dispersity ($\mathrm{SD} = 1$), and $L=32$.}\label{fig:deltavspdifferentm}
\end{figure}

An increase in the aspect ratio of the rods, i.e., increase in the value of $m$ at fixed size of the system under consideration, $L$, resulted in increase of the electrical anisotropy, $\delta$, at small and moderate values of concentration, $p$, (Fig.~\ref{fig:deltavspdifferentm}).

\section{Conclusion}\label{sec:concl}
We simulated the electrical conductivity of poorly conductive films with embedded highly conductive rod-like fillers. The fillers were allowed to intersect each other. The fillers were assumed to have different lengths according to the log-normal distribution. The mean length of the fillers was fixed, while the dispersity of length was considered as an adjusted parameter.

We established that, basically, the behavior of those systems with a dispersity of rod length was similar to the behavior of the systems with fillers of equal size. However,  the phase transition insulator-to-conductor occurred at a smaller concentration of fillers when the length dispersity increased. In the thermodynamic limit, the percolation threshold decreased linearly as the length dispersity increased. For the finite-size films, the values of the percolation thresholds along and perpendicular to the direction of rod alignment ($p_\parallel$ and $p_\perp$, respectively) were different. The effect of the system size on the electrical conductivity was significant only in the vicinity of the percolation threshold.

In films with completely aligned rods, the transversal ``intrinsic conductivity'' was insensitive to the length dispersity; by contrast, the longitudinal ``intrinsic conductivity'' increased as the dispersity increased.

The length dispersity had a more pronounced effect on the longitudinal electrical conductivity while the transversal conductivity was less sensitive to the length dispersity. The greater the extent of alignment, the more noticeable this effect. The anisotropy of the electrical conductivity was more pronounced between $p_\parallel$ and $p_\perp$.

All in all, the greater electrical conductivity corresponded to those films with completely aligned rods,  which have a high aspect ratio and large length dispersity. This fact suggests that a dispersity of filler lengths is desirable to produce films with high electrical anisotropy.

\begin{acknowledgments}
We acknowledge the funding from the Ministry of Education and Science of the Russian Federation, Project No.~3.959.2017/4.6 (Y.Y.T., I.V.V., A.V.E., V.A.G., P.G.S.) and from the National Academy of Sciences of Ukraine, Projects No.~2.16.1.6. (0117U004046), 43/19-H, and 15F (0117U006352)  (N.I.L.). 
\end{acknowledgments}

\bibliography{dispersity}

\end{document}